# SCENARIO ANALYSIS, DECISION TREES AND SIMULATION FOR COST BENEFIT ANALYSIS OF CARGO SCREENING PROCESSES

**Galina Sherman (a), Peer-Olaf Siebers (b), Uwe Aickelin (c), David Menachof (d)**

(a)(d) Business School, Hull University, Hull HU6 7RX, UK
(b)(c) Computer Science, Nottingham University, Nottingham NG8 1BB, UK

(a) G.Sherman@2008.hull.ac.uk, (b, c) pos, uxa@cs.nott.ac.uk, (d) D.Menachof@hull.ac.uk

**ABSTRACT**
In this paper we present our ideas for conducting a cost benefit analysis by using three different methods: scenario analysis, decision trees and simulation. Then we introduce our case study and examine these methods in a real world situation. We show how these tools can be used and what the results are for each of them. Our aim is to conduct a comparison of these different probabilistic methods of estimating costs for port security risk assessment studies. Methodologically, we are trying to understand the limits of all the tools mentioned above by focusing on rare events.

## 1. INTRODUCTION

The use of cost benefit analysis is described in the literature from the early 19$^{th}$ century. Already at that time this approach was used by US governmental agencies in environmental management (Hanley and Spash 1993). According to Guess and Farnham (2000), since 1960 the application of cost benefit analysis was expanded to "human beings" and "physical investment programs".

The decision of rejecting or accepting a program was made according to the benefits versus the costs of a program, if the benefits were greater than the costs or the same, the program would be accepted. In addition, the authors point out that cost benefit analysis were also used as a tool for policies evaluation and comparison, while the measure could change from benefits to efficiency, although the one who would enjoy the benefits would not be the agency that undertook the project.

Cargo screening processes are used to detect different threats such as nuclear, chemical and radiological weapons, smuggling and contraband and sometimes even stowaways. While protecting against threats the port operator's interest is to keep the port performance as smooth as possible. To screen all sea freight is a challenging task because of the huge volume of goods that pass around the globe annually. Branch (1986) argues that the amount of cargo flows in the seaports is more than 90% of total volume of worldwide cargo, which represents approximately 99% of the impact on the economy of the world.

In addition to the huge volume of cargo, very little is known about efficiency of the technology which is used. Some manufacturers have benchmarks, but the benchmarks are valid according to an experimental environment rather than to a real world. It is very rare to find unbiased benchmarks that were made according to the real world.

In this paper we examine different probabilistic techniques that are used for conducting cost benefit analysis, in particular we focus on scenario analysis, decision trees and simulation. We explore the advantages and disadvantages of using these tools, while conducting the cost benefit analysis of different cargo screening policies. The aim is a comparison of different strategies and identification of the data requirements for cost benefit analysis and how they can be used in security research.

Our main research hypothesis is that simulation is the more suitable tool than decision tree and scenario analysis in conduction of cost benefit analysis while dealing with security issues and cargo screening. In addition we compare and examine the data requirements and results between scenario analysis decision tree and simulation while implementing different search strategies. This will allow us to produce a data requirement analysis that can be used in security research.

In comparison with other techniques, such as cost effectiveness and cost utility analysis, cost benefit analysis has an advantage of comparison between wider scopes of possibilities. However, sometimes it is impossible to give a monetary value to the costs or to the benefits (Guess and Farnham 2000). Sekine et al. (2006) suggest using Pareto efficiency set to compare between different inspection policies.

Our aim is to apply these approaches in a real world situation; as a result a case study approach is chosen as a most suitable research method for the data collection. Furthermore, Farrow and Shapiro (2009) identify that the literature dealing with the cost benefit analysis is often based on a case study methodology while focusing at a sensitive topic such as security policy.

We adopt the assumption made by Kleindorfer and Saad (2005) that every business is interested in the trade off between the costs of risk mitigation investments and the expected losses of the potential disruptions.

Therefore, cost-benefit analysis provides a good tool to investigate different cargo screening scenarios. In addition, we attempt to advance the use of simulation as a powerful tool at the operational and strategic level. In combination with cost benefit analysis, we intend to give it broader use when dealing with security issues.

We investigate our hypothesis by first comparing the results of different scenarios by using scenario analysis, decision trees and simulation. We also validate our basic simulation by comparing its results with the results from the decision tree. Afterwards we compare the monetary values of different screening policies using scenario analysis, decision tree and simulation.

Our results show that simulation is more powerful and flexible than decision trees, while decision trees are more detailed than scenario analysis. Furthermore, we find scenario analysis very general and not as sensitive or flexible. However, this approach requires less data than others. Both simulation and decision trees allow us to conduct a basic what if analysis, e.g. setting the queue to a fixed maximum capacity. However, the manipulation of parameters is much more convenient in simulation. Moreover, realistic traffic patterns and queuing discipline cannot easily be replicated in decision trees.

Overall, our results and conclusions lead us to a data requirement analysis that can be used in security research. For our further research we want to apply our findings from current paper to our next case study that will take place during this year in the port of Dover. In addition, our results help to the other users to decide which technique is an appropriate for them and suitable according to the data available and to the detailed level of results they want to achieve.

This paper contains four sections as follows: Section two contains the background of previous research about risk, risk management, scenario analysis; decision trees, simulation and cost benefit analysis. In section three, we introduce our case study and the comparison of the approaches used on its basis. Section four contains our findings and conclusions.

## 2. BACKGROUND

In this section, we present the background to the methods we use. In addition, we provide a basic definition of risk. We define a risk as a threat that has a probability to occur vs. consequences (The Royal Society, 2002; Sheffi 2005). As a part of the risk management we compare different approaches that can be used for the policies comparison and express the results in the monetary values.

### 2.1. Scenario analysis

Scenario analysis is often used to analyse possible future scenarios by considering possible best, worst, and average outcomes. According to Damodaran, (2007) this technique is suitable for single events. According to Daellenbach and McNickle (2005), any business faces uncertainty and as a result creates an unlimited number of possible futures to be considered. However, the number of possible scenarios to be considered is limited to three or four in scenario analysis.

### 2.2. Decision trees

A decision tree is a decision support tool (diagram) used in operational research. It can be helpful in deciding about strategies and dealing with conditional probabilities. According to Anderson et al. (1985) decision trees are a part of the decision theory approach widely used by decision makers while dealing with few possible solutions. Decision trees are diagrams that can be used to represent decision problems so that their structure is made clearer. Unlike decision tables, decision trees can be used to represent problems involving sequences of decisions, where decisions have to be made at different stages in the problem.

### 2.3. Simulation

Simulation is widely used in the supply chain. According to Turner and Williams (2005), modelling complicated supply chain systems gives the ability to experiment with different scenarios and has the power of generalization of the insight on the performance of the complicated systems.

In addition, Wilson (2005) also confirms the usefulness of simulation for this and states that "Simulation modeling allows the analysis or prediction of operational effectiveness, efficiency, and detection rates (performance) of existing or proposed security systems under different configurations or operating policies before the existing systems are actually changed or a new system is built, eliminating the risk of unforeseen bottlenecks, under- or over-utilization of resources, or failure to meet specified security system requirements". The author suggests making assumptions in order to simplify the model, but on the other hand he suggests achieving flexibility in the model.

### 2.4. Cost benefit analysis

As we already have mentioned in the introduction part, cost benefit analysis is an approach that has a long implication history. In economic literature, it is compared with cost effectiveness and cost utility analysis. The difference between these three approaches that the cost benefit analysis allows a comparison of wider range of scenarios, because the costs and the results have a monetary expression unlike two other techniques that are observed by using a single result every time.

In the medical literature, we find a successful use of simulation for cost benefit analysis. For example, Habbema et al. (1987) suggest simulation as an appropriate technique to conduct the cost benefit analysis to compare two different screening policies, where "screening" refers to cancer screening. The authors use a micro simulation approach to explore different scenarios and their outputs. Similar to Habbema et al. (1987) Pilgrim et al. (2009) suggest conducting a cost effectiveness analysis for cancer screening policies while using discrete event simulation. The authors support their choice of methodology with previous research using the same research strategy.

Jacobson et al. (2006) suggest the cost benefit analysis in evaluation of cost effectiveness while screening 100% of the cargo and using single or double devices. The authors propose a model to measure costs versus benefits while using different configurations of explosive detection systems and explosive trace detection machines in US airports. They suggest a cost model that contains direct costs and indirect costs associated with costs of false clears, this cost is based on subjective probability to occur. After introducing the model the authors assume that this indirect cost is large and very difficult to estimate and propose a model without it.

In the next section we present our research question and afterwards the case study, on its basis we will examine the scenario analysis decision tree and simulation approaches for the cost benefit analysis.

## 3. CASE STUDY

In order to achieve our research aim (comparing scenario analysis, decision trees and simulation for a cost benefit analysis) we chose a case study approach as our research methodology. Our first case study involves the cargo screening facilities of the ferry port of Calais (France). We conduct this case study in collaboration with the UK Border Agency (UKBA). Calais was chosen for the following reasons: the limited number of links – Calais operates only with Dover leading to a simple cargo flow and there is only one major threat of interest to the British government (clandestines). Clandestines are people who are trying to enter the UK illegally – i.e. without having proper papers and documents.

In Calais there are two security areas, one is operated by French authorities and the other one is operated by UK Border Agency. According to the data that we have collected from the field between April 2007 and April 2008 about 900,000 lorries passed the border and approximately 0.4% of the lorries had additional human freight (UKBA 2008).

For our case study, we have conducted scenario analysis, built a decision tree that fully represents the flow inside our system and have built a simulation model of the system using the same input data as in the decision tree. Comparing and contrasting results of both, decision tree and simulation, allowed us to validate the models.

While we used Microsoft Excel and simple spreadsheet calculation for modelling our decision trees our Monte Carlo and then discrete event simulation was built using AnyLogic software (XJ Technologies 2010). The process flow representation in the simulation is equivalent to the decision tree layout. However, the simulation uses probabilities and frequency distributions, e.g. exponential arrival times of lorries. This randomness (e.g. slightly different number of arrivals each time) requires us to undertake several replications of each simulation scenario and calculate the means of the simulation model outputs. These means we compare with the results of the decision tree to validate our model. Then we conduct a range of experiments by using all three approaches, we translate the results to the monetary value and compare the outcomes.

We then extend the simulation by changing arrival times from simple exponential distributions to more realistic traffic patterns such as peak hours and seasonal patterns. In addition, we modify our simulation model by adding maximum queue capacities various stations in the system. If a queue reaches its capacity, lorries will bypass the station without checks. These modifications reflect the real world and make our simulation model more realistic than the decision tree (which then has the ability to change the results).

### 3.1. Common data used for all the approaches
The common data that we use for all three scenarios is summarised in the next table:

Table 1: Common Data

| | |
|---|---|
| Total number of lorries | 900,000 |
| Cost per missed positive lorry | £400,000 |
| Cost to increase searches by 0% | £0 |
| Cost to increase searches by 10% | £5,000,000 |
| Cost to increase searches by 20% | £10,000,000 |
| *Current number of positive lorries found* | |
| France (end1) | 1,800 |
| UK Shed (end2) | 890 |
| UK Berth (end3) | 784 |
| UK total (end2 + end3) | 1,674 |
| Estimate of positive lorries not detected | 150 |

We consider two factors with three scenarios each in our scenario analysis: Traffic growth and clandestine growth (Table 2). For each factor and scenario combination, we have estimated the probability of it happening, as described in the following paragraphs. The question we are trying to answer is how the UK Border Agency (UKBA) should respond to these scenarios. We assume that there are three possible responses: increasing the searches by either 0%, 10% or 20%.

Traffic growth represents the percentage increase in lorry traffic that passes through Calais. Exact forecasts for traffic growth vary, but Calais port is already planning to open a second terminal by around 2020. By then roughly a doubling of traffic is expected, i.e. +100% in ten years. Thus an annual traffic growth of 0%-20% seems a realistic factor range, with an increase of 10% most likely. It is assumed that any increase in traffic is proportional, i.e. the ratio of soft to hard sided lorries remains the same.

The second factor under consideration is clandestine growth. This is the most unpredictable of the three factors, as clandestine numbers greatly vary from year to year based largely on external factors such as the economic attractiveness of the UK, the number and intensity of wars and other conflicts worldwide and other political initiatives.

Local aspects also play a role, for example an increase in searches in Calais can displace clandestines to other nearby ports and vice versa. Due to the uncertainty attached to this factor, a range of +25% to -50% is considered, with all scenarios being equally likely. A higher maximum decrease than increase is assumed to the recent clearing in late 2009 of the Calais "jungle" (illegal encampment of clandestines near the port). We will assume in the following that any changes in clandestine numbers will proportionally effect successful and unsuccessful clandestines.

Table 2: two factors with three scenarios and one decision variable with three options.

| Traffic Growth (TG) | p (TG) | Clandestine Growth (CG) | p (CG) | Search Growth (SG) |
|---|---|---|---|---|
| +0% | 0.25 | -50% | 0.33 | +0% |
| +10% | 0.5 | +0% | 0.33 | +10% |
| +20% | 0.25 | +25% | 0.33 | +20% |

Combining the above information, we arrive at the following combined probabilities of each scenario to occur:

Table 3: combined probabilities assuming independence of probabilities

|  | -50% CG | 0% CG | +25% CG |
|---|---|---|---|
| 0% TG | 0.083 | 0.083 | 0.083 |
| 10% TG | 0.167 | 0.167 | 0.167 |
| 20% TG | 0.083 | 0.083 | 0.083 |

It is estimated by the UKBA that each clandestine that reaches the UK costs the government approximately £20,000 per year. Moreover, it is estimated that he average duration of a stay of a clandestine in the UK is five years, so the total cost of each clandestine slipping through the search in Calais is £100,000.

The cost for increasing the search capacity in Calais is more difficult to estimate, as there is a mixture of fixed and variable cost and operations are often jointly performed by French, British and private contractors. However, if we concentrate on UKBA's costs, we can arrive at some reasonable estimates, if we assume that any increase in searches would result in a percentage increase in staff and infrastructure cost. Thus we estimate that a 10% increase in search activity (10% SG) would cost £5M and a 20% increase £10M (20% SG).

Search growth describes the percentage increase in search activity by the UK Border Agency (UKBA). Currently, UKBA searches 33% of traffic. To keep this proportion stable, UKBA will need to respond to a growth in traffic by increasing the number of lorries it searches. At the same time, there is political pressure to search more vehicles, whilst budget pressures limit the number of vehicles that can be inspected. Thus we assume that search growth may also vary between 0% and 20%. As before, we assume that any increase in search activity is proportional to hard and soft sided lorries.

As the status quo is 33% of vehicles are searched by UKBA, we can calculate resultant percentages of vehicles searched by combining the above two factors (assuming linear relationships). For example, if traffic increase is matched by search increase, this will remain the same. Or if there is a +10% traffic growth and +0% search growth results in 33% * (100%/110%) = 30% of vehicles searched.

A key question is the relationship between the percentage of vehicles searched versus the number of clandestines found or more importantly, the number of clandestines not found. At a search intensity of 33%, UKBA finds approximately 1,674 lorries in Calais with additional cargo. A best estimate of "successful" clandestines is approximately 50 per month (600 per year) or 150 lorries per year. Establishing a clear relationship between these 150 and the figure of 1,674 is difficult, as 1,674 lorries does not represent unique attempts by the clandestines. Unsuccessful clandestines will try time after time.

It is probably a fair assumption that an increase in searches will yield a decrease in the number of successful clandestines and vice versa. In absence of further information and considering that the variation of percentage of searches is in a relatively limited range of 27.5% to 39.6%, we will make the same assumption here as in the rest of the scenario analysis: the relationship between both parameters is linear. Based on this, we obtain the number of clandestines missed as given in Table 2, e.g. searching only 30% of traffic results in 600* (33%/30%) = 660 missed clandestines.

### 3.2. Calais Scenario Analysis

We use different methods for estimating the 'adjusted' number of positive lorries found if there is no growth of positive lorries. Once we have the matrix (iteration over our two factors) we conduct some data analysis to estimate the costs. The data analysis is the same for all the different approaches we will present.

We demonstrate this for scenario analysis and for all others we will only report on the key outputs (adjusted number of positive lorries found for CG=0, total expected costs). Finally, we will compare the expected costs that resulted from the different models.

Table 4- Proportion of vehicles searched

| TG vs. SG | SG 0% | SG +10% | SG +20% |
|---|---|---|---|
| TG 0% | 0.3300 | 0.3630 | 0.3960 |
| TG 10% | 0.3000 | 0.3300 | 0.3600 |
| TG 20% | 0.2750 | 0.3025 | 0.3300 |

Table 5 is the result from the scenario analysis: number of positive lorries found. The cost estimation that follows below is the same for all methods discussed in this paper and will only be shown once in detail.

Table 5: Adjusted number of positive lorries found if CG = 0%. Calculated based on Table 4 where 0.3300 represents 1,674 lorries.

| TG vs. SG | SG 0% | SG +10% | SG +20% |
|---|---|---|---|
| TG 0% | 1674.0 | 1841.4 | 2008.8 |
| TG 10% | 1521.8 | 1674.0 | 1826.2 |
| TG 20% | 1395.0 | 1534.5 | 1674.0 |

Table 6: Relative number of positive lorries found when compared to base scenario if CG = 0%

| TG vs. SG | SG 0% | SG +10% | SG +20% |
|---|---|---|---|
| TG 0% | 1 | 1.1 | 1.2 |
| TG 10% | 0.909091 | 1 | 1.090909 |
| TG 20% | 0.833333 | 0.916667 | 1 |

Table 7: Number of positive lorries missed if CG = 0%. Calculated based on the probabilities in Table 4, where 0.3300 results in 150 missed positive lorries. Similar Tables can be computed for the other CG values.

| TG vs. SG | SG 0% | SG +10% | SG +20% |
|---|---|---|---|
| TG 0% | 150.0 | 136.4 | 125.0 |
| TG 10% | 165.0 | 150.0 | 137.5 |
| TG 20% | 180.0 | 163.6 | 150.0 |

Table 8: Cost of extra searches – as mentioned before.

| TG vs. SG | SG 0% | SG +10% | SG +20% |
|---|---|---|---|
| TG 0% | £0 | £5,000,000 | £10,000,000 |
| TG 10% | £0 | £5,000,000 | £10,000,000 |
| TG 20% | £0 | £5,000,000 | £10,000,000 |

Table 9: Relative number of positive lorries missed compared to the base scenario (inverse of Table 6)

| TG vs. SG | SG 0% | SG +10% | SG +20% |
|---|---|---|---|
| TG 0% | 1.00 | 0.91 | 0.83 |
| TG 10% | 1.10 | 1.00 | 0.92 |
| TG 20% | 1.20 | 1.09 | 1.00 |

Table 10: Expected costs excluding SG costs for CG = 0%. Calculated by combining the information of table 8 with that of Table 9, where 1.00 means 150 lorries at a cost of £400,000 each.

| TG vs. SG | SG 0% | SG +10% | SG +20% |
|---|---|---|---|
| TG 0% | £60,000,000 | £59,545,455 | £60,000,000 |
| TG 10% | £66,000,000 | £65,000,000 | £65,000,000 |
| TG 20% | £72,000,000 | £70,454,545 | £70,000,000 |

Table 10 can be used to calculate the tables for CG=-50% and CG=25% by multiplying each value in the table with 1+CG(x) where CG(x) is the proportion of Clandestine Growth.

Overall, we can conclude from the scenario analysis that the best option seems to be to change SG by 10%. In our decision tree and simulation we use the same assumptions, same costs and same scenarios as described above. However, when we use additional data for an approach we will state it in the relevant section.

### 3.3. Calais Decision Tree

In order to conduct a comparison between different techniques we will use for the decision tree the same scenarios that we have used for the scenario analysis. Also, here we will consider the same two factors: traffic growth and clandestine growth and as a reaction search growth. We investigate what is an appropriate policy that UKBA should adopt according to the decision tree. In addition we want to know the difference in the data inputs and outputs between these approaches.

For our case study we have built a decision tree that fully represents the flow inside our system (Figure 1). Building both allows us to check the results and validate the models as the results should be identical.

Figure 1: Decision Tree

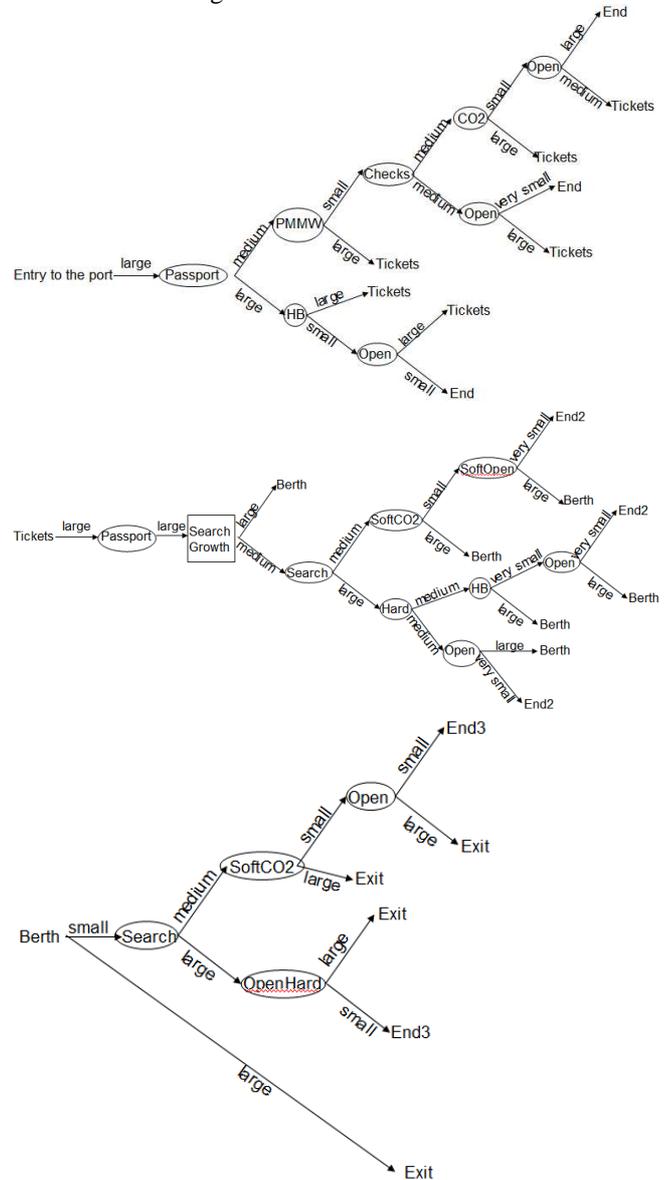

Building the decision tree using probabilities demands more data on the one hand; however it allows us to receive more precise outcomes on the other hand. Due to the sensitivity of the data we change the numerical probabilities in the decision tree to their equivalent in words (see Table 1).

Table 11: probabilities that are used in the decision tree and their equivalent in words.

| Probabilities | p>50% | 10%<p≤50% | 1%<p≤10% | p≤1% |
|---|---|---|---|---|
| Equivalence | large | medium | small | very small |

From the decision tree we can calculate the following results, which are identical to the scenario analysis, apart from small rounding errors (due to more complex spreadsheet calculations).

Table 12: Decision Tree results: Number of positive lorries found if CG = 0%.

| TG vs. SG | SG 0% | SG +10% | SG +20% |
|---|---|---|---|
| TG 0% | 1674 | 1841 | 2008 |
| TG 10% | 1522 | 1674 | 1826 |
| TG 20% | 1395 | 1534 | 1674 |

Using the same combined probabilities for each scenario as we have used in scenario analysis we find that the results between two approaches are almost identical in the monetary value and according to both of them we should adopt the same search strategy. Overall we conclude from the decision tree that the best option seems to be to change SG by 10%.

### 3.4. Calais Simulation

Besides the data already mentioned we have collected data on operation times of the activities in the shed and the berth and from ferry operation manuals. We have used the decision tree graphical representation as a conceptual model basis for the Monte Carlo Simulation and Discrete Event Simulation implementation.

Figure 2: The French site (AnyLogic Simulation)

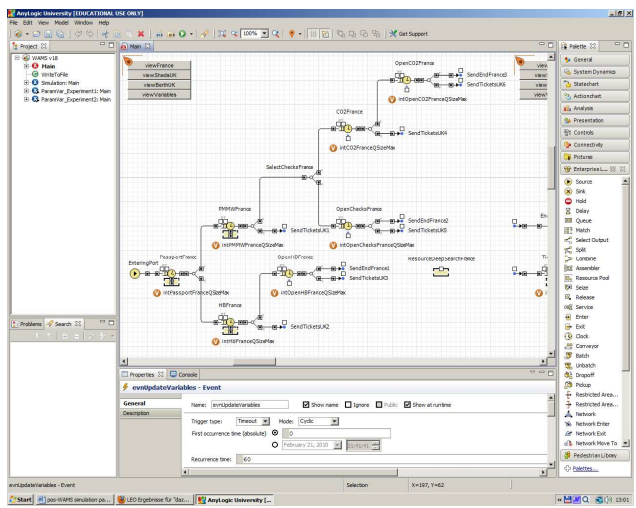

We use the same simulation model for both, Monte Carlo Simulation and Discrete Event Simulation. The parameter set loaded in the initialisation phase of the simulation determines the type of simulation that will be run.

The first parameter set sets all delay times to zero and all queue sizes to 1,000,000. This allows emulating a Monte Carlo Simulation where all events are executed in the correct sequence but the time it takes to execute them is zero.

The second parameter set defines triangular frequency distributions for the delay times (based on our collected case study data), sets all queue sizes to 1,000,000, and defines the resources that are available (based on our collected case study data). Queue sizes can later be restricted by defining routing rules.

For both, Monte Carlo Simulation and Discrete Event Simulation, routing decision (e.g. which lorries to inspect and how to inspect them) are derived from uniform probability distributions (given probabilities are based on our collected case study data).

Having stochastic inputs means that we also have stochastic outputs. Therefore, we have to do multiple runs. We have conducted some tests to determine the number of replication required using confidence intervals (95%) following the guidelines in Robinson (2004). The test result suggests to run at least 8 replications. To be on the safe side we decided to do 10 replication for each iteration of our experiments. To report our experimental results we use mean value as a point estimator and standard deviation as an estimator of the variability of the results.

### 3.5. Monte Carlo Simulation

We have run all scenarios as defined in the previous sections. The results are as follows:

Table 13: Monte Carlo Simulation results: Number of positive lorries found if CG = 0%.

| TG vs. SG | SG 0% | SG +10% | SG +20% |
|---|---|---|---|
| TG 0% | 1678.75 | 1846.25 | 2027.75 |
| TG 10% | 1531.30 | 1674.15 | 1827.50 |
| TG 20% | 1404.25 | 1540.90 | 1670.70 |

Comparing the results with the ones from decision trees shows the following:

Table 14: Comparing Decision Tree Results with Monte Carlo Simulation results (errors)

| SG 0% | SG +10% | SG +20% |
|---|---|---|
| -5.0 | -5.2 | -19.4 |
| -9.6 | -0.4 | -1.7 |
| -9.4 | -6.6 | 3.0 |

We find that the differences are relatively small and can be attributed to the fact that we use a stochastic method. There are different types of simulation. In this paper we look at static discrete stochastic simulation (Monte Carlo Simulation) and dynamic discrete stochastic simulation (Discrete Event Simulation. The difference between these simulation approaches is that in Monte Carlo Simulation time does not play natural role but in Discrete Event Simulation it is (Kelton et al., 2010). While the first is often used in risk assessment the second is often used when further investigation into the system behaviour on the operational level are required for the decision making.

Therefore, for our comparison of probabilistic risk analysis methods we use a Monte Carlo Simulation model. We then extend this Monte Carlo Simulation model to a Discrete Event Simulation model by adding elements that are linked to time (e.g. arrival rates, delays, queues) and constraints that are linked to time (e.g. queue size restrictions) and new performance measures linked to time (e.g. utilisation, time in system, max queue length) in order to demonstrate the full power of simulation as a tool in the decision making process.

As we mentioned before we use the same combined probabilities for each scenario as we have used in scenario analysis and decision tree. We find that the simulation results are similar to the other two approaches. Although here, according to the simulation outcomes we should adopt the same search strategy, e.g. SG 10%. The difference in the monetary value between simulation and other two approaches is very as result of calculating the costs according to the averages (decision tree and scenario analysis) vs. according to the output for the simulation.

### 3.6. Discrete Event Simulation – DES 0

In aim to upgrade our model to Discrete Event Simulation we have added additional data that represents the real world situation. The extra data that we use is service times and resources we run our simulation again to check if this additional data makes an effect on the results. We find that the impact is very small and as a result we would choose the same strategy as before SG 10%.

Table 15: Standard Discrete Event Simulation Results: Number of positive lorries found if CG = 0%.

| TG vs. SG | SG 0% | SG +10% | SG +20% |
|---|---|---|---|
| TG 0% | 1674.50 | 1833.00 | 2013.90 |
| TG 10% | 1512.00 | 1667.20 | 1818.15 |
| TG 20% | 1387.65 | 1527.90 | 1694.20 |

To show the additional possibilities of Discrete Event Simulation we have added some additional features we want to demonstrate here: These are (1) variable arrival rates, (2) queue size restrictions at sheds UK and (3) combination of first two features.

#### 3.6.1. Variable arrival rates – DES 1

After running the Discrete Event Simulation with variable arrival rates we find that also here the results are very similar to previous runs and scenarios and the appropriate strategy to adopt is SG 10%. The monetary difference is very small. The results are logical if there is no queue capacity involved and the only change is at the arrival rates.

Table 16: Discrete Event Simulation with Variable Arrival Rates results: Number of positive lorries found if CG = 0%.

| TG vs. SG | SG 0% | SG +10% | SG +20% |
|---|---|---|---|
| TG 0% | 1681.55 | 1843.00 | 2008.70 |
| TG 10% | 1519.25 | 1687.20 | 1852.85 |
| TG 20% | 1385.05 | 1534.85 | 1658.85 |

#### 3.6.2. Queue size restrictions at sheds UK – DES 2

At this stage we run the Discrete Event Simulation while queue capacity is set, we find that here the results are different than in previous runs and scenarios, but the appropriate strategy to adopt is still SG 10%. The monetary difference is bigger. Although, we find that these results are logical the queue capacity only has its impact on the system but this effect is not big enough to make change in the strategy.

Table 17: Discrete Event Simulation with Queue Size Restrictions results: Number of positive lorries found if CG = 0%.

| TG vs. SG | SG 0% | SG +10% | SG +20% |
|---|---|---|---|
| TG 0% | 1666.95 | 1833.30 | 1998.90 |
| TG 10% | 1507.80 | 1657.85 | 1795.35 |
| TG 20% | 1383.90 | 1526.70 | 1653.95 |

#### 3.6.3. Variable arrival rates and queue size restrictions at sheds UK – DES 3

As a final stage for this paper we have combined the two previous scenarios for Discrete Event Simulation: the arrival rate and queue capacity. From the results at this stage we find that the most appropriate strategy to adopt is to keep SG 0%. The reason for the different outcome lies in the combination of arrival rates and queue capacity, when the effort to search more is wasted by queue jumping during the peak times.

Table 18: Discrete Event Simulation with Variable Arrival Rates and Queue Size Restrictions results: Number of positive lorries found if CG = 0%.

| TG vs. SG | SG 0% | SG +10% | SG +20% |
|---|---|---|---|
| TG 0% | 1626.35 | 1788.10 | 1891.50 |
| TG 10% | 1500.75 | 1619.15 | 1731.25 |
| TG 20% | 1374.8 | 1492 | 1600.9 |

### 4. COMPARISON AND CONCLUSIONS

We managed to gain similar results for all methods we looked at, e.g. scenario analysis, decision tree and simulation (Monte-Carlo). The simulation (DE) results are different from the other approaches and as we explained in the relevant section are logical.

Table 19: Overall cost comparisons of all methodologies.

|        | Total expected costs |            |            | Cheapest option |
|--------|-----------|-----------|-----------|---|
| Option | 1: SG=0%  | 2: SG=10% | 3: SG=20% |   |
| SA     | £60,500,000 | £60,000,000 | £60,416,667 | 2 |
| DT     | £60,497,446 | £60,000,000 | £60,418,795 | 2 |
| MCS    | £60,335,818 | £60,058,184 | £60,461,341 | 2 |
| DES 0  | £60,797,873 | £60,250,740 | £60,350,102 | 2 |
| DES 1  | £60,881,284 | £60,017,602 | £60,406,308 | 2 |
| DES 2  | £60,714,953 | £60,166,442 | £60,857,915 | 2 |
| DES 3  | £59,817,382 | £60,116,618 | £61,624,835 | 1 |

Table 20: Relative cost comparisons of all methodologies.

|        | Relative difference from lowest costs | | | Cheapest option |
|--------|-----------|-----------|-----------|---|
| Option | 1: SG=0%  | 2: SG=10% | 3: SG=20% |   |
| SA     | £500,000  | £0        | £416,667  | 2 |
| DT     | £497,446  | £0        | £418,795  | 2 |
| MCS    | £277,633  | £0        | £403,156  | 2 |
| DES 0  | £547,133  | £0        | £99,362   | 2 |
| DES 1  | £863,682  | £0        | £388,706  | 2 |
| DES 2  | £548,511  | £0        | £691,473  | 2 |
| DES 3  | £0        | £299,236  | £1,807,453 | 1 |

However, to advice which technology to use when and what is involved regarding factors like data requirement etc. we suggest to use the next table:

Table 21: Factors to take into consideration before making decisions (SA – Scenario Analysis, DT – Decision Tree, MC – Monte Carlo Simulation, DES – Discrete Event Simulation)

|                |                           | SA | DT | MC | DES |
|----------------|---------------------------|----|----|----|-----|
| Risk type      | Discrete / Continuous     | D  | D  | C  | C   |
|                | Correlated / Independent  | C  | I  | both | both |
|                | Sequential / Concurrent   | C  | S  | both | both |
| Decision process | Strategic / Operational | S  | S  | S  | O   |
|                | Broad / Detailed          | B  | B  | B  | D   |
| Model Characteristics: Low, Medium, High | Complexity | L | M | H | H |
|                | Data requirements         | L  | L  | M  | H   |
|                | Tool costs                | L  | L  | M  | H   |
|                | Training costs            | L  | L  | H  | H   |
|                | Assumptions               | H  | M  | L  | L   |

While dealing with continuous types of risks the simulation approach will be more suitable and it does not depend if the risk type is correlated or independent, sequential or concurrent. However, for scenario analysis and decision tree the situation is different. The scenario analysis can be used for correlated, concurrent and discrete type of risk while decision trees can be used for independent, sequential and discrete types.

The decision process level and the information required for our three basic approaches will be similarly applied at the strategic level and require broad information. On the other hand Discrete Event Simulation requires detailed information and can be applied on the operational level.

Total costs of the approaches (including training costs) vary from low for scenario analysis (that require just pen and paper) and decision trees to high for simulation.

Our future plans are to get away from probabilistic routing to factual routing. This means abandoning probabilities at the routing decision points and making the decision based on the attributes of the object that is at the decision point. The attribute would be clandestine on board, and the correct decision would be made based on a probability. This allows us to integrate (and test) different detection rates for different sensors.

**ACKNOWLEDGMENTS**
This project is supported by the EPSRC, grant number EP/G004234/1 and the UK Border Agency.

**AUTHORS BIOGRAPHY**

GALINA SHERMAN is a PhD student at Hull University, Business School. Her current research is related to supply chain management, risk analysis and rare event modelling.

PEER-OLAF SIEBERS is a Research Fellow at The University of Nottingham, School of Computer Science. His main research interest is the application of computer simulation to study human-centric complex adaptive systems. This is a highly interdisciplinary research field, involving disciplines like social science, psychology, management science, operations research, economics and engineering. For more information see http://www.cs.nott.ac.uk/~pos/

UWE AICKELIN is Professor of Computer Science and an Advanced EPSRC Research Fellow at The University of Nottingham, School of Computer Science. His main research interests are mathematical modelling, agent-based simulation, heuristic optimisation and artificial immune systems. For more information see http://www.cs.nott.ac.uk/~uxa/

DAVID MENACHOF is Professor of Port Logistics at the Business School, University of Hull. He received his Doctorate from the University of Tennessee and was the recipient of the Council of Logistics Management's Doctoral Dissertation Award in 1993. He is a Fulbright Scholar, having spent a year in Odessa, Ukraine as expert in logistics and distribution. His work has been published and presented in journals and conferences around the world. His research interests include supply chain security and risk, global supply chain issues, liner shipping and containerisation and financial techniques applicable to logistics.